 \documentstyle[preprint,eqsecnum,aps]{revtex}

\begin{document}
\draft
\preprint{HEP/123-qed}

\title{Charging effects in the ac conductance of a double barrier resonant
tunneling structure}

\author{M.P. Anantram}
\address{ NASA Ames Research Center,
Mail Stop: T27-A,
Moffett Field,
CA 94035-1000, U.S.A}

%\date{\today}

\maketitle

\begin{abstract}
There have been many studies of the linear response ac conductance of
a double barrier resonant tunneling structure (DBRTS). While these
studies are important, they fail to self-consistently include the
effect of time dependent charge density in the well. In this paper,
we calculate the ac conductance by including the effect of time 
dependent charge density in the well in a self-consistent manner. The
charge density in the well contributes to both the flow of displacement
currents and the time dependent potential in the well. We find that 
including these effects can make a significant difference to the ac 
conductance and the total ac current is not equal to the average of 
non-selfconsitently calculated conduction currents in the two contacts,
an often made assumption. This is illustrated by comparing the results
obtained with and without the effect of the time dependent charge 
density included properly. 
\end{abstract}

\vspace{0.25in}

\bf{Appeared in J. Phys: Condens. Matter vol. 10, p. 9015 (1998)}

\vspace{0.25in}

\pacs{ }

%}

\section{\bf INTRODUCTION}

Double-barrier resonant tunneling structures (DBRTS) have been of
interest because of possible device applications in building logic 
circuits, oscillators, detectors etc., and it has much to offer in
the study of physics of confined structures. The dc charactersitics
have been studied extensively by including effects of charging and 
inelastic scattering. Reference \cite{Buot93a} offers a comprehensive
review of applications and basic physics of DBRTS. In comparison, there
are only a few studies of ac response over various frequencies regimes 
\onlinecite{Frensley88,Buot93b,Jacoboni90,Liu91,Chen91,Fu93}. While 
some of these studies are based on simulating a realistic device using
detailed numerical procedures~\cite{Frensley88,Buot93b}, 
others are based on simple models~\cite{Jacoboni90,Liu91,Chen91,Fu93}.
These calculations however do not include the effect of time
varying charge density in the well, which is important in determining
ac conductance\onlinecite{Buot93b,Jacoboni90,Landauer3,Buttiker2,Buttiker3,Buttiker4}.
Reference \onlinecite{Landauer3} has discussed the pit falls of many 
existing ac conductance theories qualitatively and reference 
\onlinecite{Buttiker2} formulated the theory of ac conductance in the
linear response and low frequency regime as applicable to  mesoscopic
structures by including effects of charging. Subsequently, reference 
\cite{Anantram1} used a non-equilibrium greens function approach to 
provide a formulation that can be used at finite biases and large 
frequencies including effects of charging in the well and phonon 
scattering \cite{Christen1}.  The effect of time dependent charge 
density in the well is of importance in determining the dynamics 
because electrons in the well image to the outside world, which 
includes the contacts. This causes flow of displacement currents and 
contributes to the ac potential in the well. The role of these factors
in determining the ac conductance of a DBRTS is not clear from previous
work. In this paper, we study the ac conductance of a DBRTS with the 
aim of illustrating the role of imaging of well-charge to contacts via
a simple model. Imaging of well-charge to the two contacts is modeled
by capacitances denoted by $C_1$ and $C_2$ (figure 1).  We would like
to clarify at the outset that the purpose of this study is to 
illustrate the importance of imaging and not to model a typical 
resonant tunneling device whose structure is considerably more 
complicated.

We derive useful expressions for the ac conductance and show that
the ac conductance depends significantly on both (i) the ratio of 
capacitances $\frac{C_1}{C}$ and $\frac{C_2}{C}$, where $C = C_1 + 
C_2$ and (ii) the value of the total capacitance between well and
contacts. The first feature follows because a time dependent 
well-charge $q(t)$ contributes to flow of displacement currents equal
to $\frac{C_1}{C} \frac{dq(t)}{dt}$ and $\frac{C_2}{C} 
\frac{dq(t)}{dt}$ in contacts 1 and 2 respectively (figure 1), where 
$q(t)$ is charge in the well at time $t$. The second feature can be 
understood by noting that time dependent charge density in the well 
contributes to ac potential of the well via a term $\frac{q(t)}{C}$ 
(figure 1). This affects the current because the ac potential in the 
well plays a role in determining both the conduction and displacement
currents. Note that there is nothing quantum mechanical about these 
two features and they would be true even if $C_1$ and $C_2$ are leaky
capacitors in a classical circuit. Quantum mechanics plays a role only
in determining the values of $q(t)$ and current.

Some previous papers calculated the ac conductance of a DBRTS by the 
following procedure (see section \ref{sect:charging}). The conduction 
currents across the two barriers are calculated by neglecting the 
contribution to ac potential in the well from the time dependent charge
density. Then the total ac current is taken to be the average of the 
conduction currents flowing across the two barriers. {\it We find such
a procedure to be valid only when the capacitances are symmetrical 
($C_1 = C_2$) and the value of $C$ is large.} 

The remainder of the paper is arranged as follows. In section 
\ref{sect:model}, we explain the model adopted in detail. In section 
\ref{sect:charging}, we discuss the effect of charging on the ac 
conductance using expressions derived in section \ref{subsect:ac-cond}.
In section \ref{subsect:num_1_mode}, we discuss the effect of charging
on the ac conductance using numerical examples. The effect of including
asymmetries in the barrier strengths and the capacitances of an 
otherwise symmetric structure are systematically studied here. We 
present our conclusions in section \ref{sect:conclusions}.

\section{{\bf MODEL}}
\label{sect:model}

\subsection{\bf Hamiltonian}
\label{subsect:hamiltonian}

We model the resonant tunneling structure by using a tight binding 
Hamiltonian~\cite{Jacoboni90,Liu91,Chen91,Fu93,Sheard1,Datta4}. The 
well is represented by a single node with a resonant level at energy 
$\epsilon_r$ and is coupled to contacts 1 and 2 (figure 1). The 
Hamiltonian of the structure is,
\begin{eqnarray}
H &=& H_D + H_C + H_{CD}        \label{eq:hamiltonian}\\
\mbox{where, }  H_D &=& \epsilon_r(t) d^{\dagger} d
\nonumber \\
H_C &=& \sum_{i,\; \alpha \in 1, \;2} ( \epsilon_{i \alpha} + e
v^{ac}_1 cos(\omega t) \delta_{1,\alpha} ) c^{\dagger}_{i \alpha}
c_{i \alpha} + ( w c^{\dagger}_{i \alpha} c_{i+1 \alpha} +c.c ) 
\nonumber \\
H_{CD} &=& \sum_{\alpha \in 1, \;2} w_{\alpha} 
c^{\dagger}_{0 \alpha} d   +   c.c  \mbox{ .}   \nonumber
\end{eqnarray}
We assume that the ac potential is applied only to contact 1. $d$ 
($d^\dagger$) and $c_{i\alpha}$ ($c_{i\alpha}^\dagger$) are 
annihilation (creation) operators for electrons in the well and
various lattice sites of contact $\alpha$ respectively. The sites
in a contact are labeled starting from $0$ which represents the 
lattice site immediately neighboring the well.
$w_1$ and $w_2$ represent the coupling between the well and site 0 of
contacts 1 and 2 respectively. $w$ represents the coupling between 
nearest neighbors in the tight binding lattice of the contacts. 
$H_D$, $H_C$ and $H_{CD}$ represent the Hamiltonians of the isolated 
well, contacts and coupling between the well and
contacts respectively. In the presence of a
dc bias $V_\alpha^{dc}$ applied to contact $\alpha$, the
on-site potential in  contact $\alpha$ increases by the dc 
bias ($\epsilon_{i\alpha} \rightarrow \epsilon_{i \alpha} + e 
V_\alpha^{dc}$).  The expression for $\epsilon_r (t)$ is,
\begin{eqnarray}
\epsilon_r (t) &=& \epsilon_{r0} + \beta_2 e V_{dc} + 
\alpha_2 e v^{ac}_1 (t) + e V_{w}^{Q} (t) \mbox{,}
\label{eq:e_r}
\end{eqnarray}
where, $\epsilon_{r0}$ is the energy of the resonant level at zero
dc and ac biases. $V_{dc}$ and $v^{ac}_1$ are dc and ac biases applied
to contact 1. $\beta_2$ and $\alpha_2$ are fractional drops of the 
external dc and ac potentials between contact 2 and the well 
respectively, in absence of charge in the well.  The last term of 
equation (\ref{eq:e_r}) represents the effect of imaging of charge in
the well. The electrons in the well image to the contacts and this is
modeled by capacitances $C_1$ and $C_2$ (figure 1), which are assumed
to be known parameters~\cite{Buttiker2,Sheard1}. 
As a consequence of imaging of well charge, the band bottom in the 
well (and hence the resonant level) changes by $V_w^Q (t)$,
\begin{eqnarray}
&& V_{w}^{Q} (t) = \frac{ q(t)}{C} = V^{dc}_{w} + v^{ac}_{w}(t),
  \label{eq:V_w} \\
&& \mbox{where,  } \;\; V^{dc}_{w} = \frac{q_{dc}(V)}{C} \;\;
\mbox{ and } \;\; v^{ac}_{w}(t) = \frac{ q(t) - q_{dc}(V) }{C}
\mbox{.}
\end{eqnarray}
$q_{dc}(V)$ represents the dc charge in the well when the applied
dc voltage is V.
$q(t)$ is the total charge in the well and $q(t) - q_{dc}(V)$ is
the time dependent component.
The total capacitance between the well and contacts is $C$ 
($= C_1 + C_2$).

The total potential of the well $V_w(t)$ is,
\begin{eqnarray}
V_w (t) = V_{w}^{Q} (t) + \beta_2  V_{dc} + \alpha_2 v^{ac}_1(t) 
\label{eq:V_w1}  \mbox{.} 
\end{eqnarray}
The first, second and third terms represent the potentials due to 
imaging of well charge, the externally applied dc and ac 
potentials respectively.

\subsection{\bf ac Conductance}
\label{subsect:ac-cond}

The applied ac voltage results in a time dependent well charge that 
images to the contacts. This (i) causes a flow of displacement 
currents and (ii) contributes to the time dependent potential of the 
well ($V^Q_w$ of equation (\ref{eq:V_w})). (ii) plays a role in 
determining the correct conduction currents. The conduction current is
in turn related to the time varying well charge by continuity equation.

In the following discussion, we first define a conductance matrix,
$g$. The conduction and displacement currents are then expressed in
terms of the elements of the conductance matrix. The procedure to 
evaluate the conductance matrix is discussed in the next subsection.

\noindent
{\bf The conductance matrix ($g$):} 
The conductance matrix element $g_{\alpha\beta}$ represents the ratio
of conduction current ($i_\alpha^c$) flowing in contact $\alpha$
as a result of an ac voltage ($v_\beta$) applied to contact $\beta$,
with the ac potential in well and contacts set equal to zero:
\begin{eqnarray}
g_{\alpha \beta}(\omega) = \frac{i_\alpha^{c}(\omega)}{v^{ac}_{\beta}
(\omega)}  \;\; \mbox{, where} \; \alpha,\; \beta \in 1, \; 2\mbox{.}
 \label{eq:g}
\end{eqnarray}
The dc voltage is set to its steady state value, i.e., the dc
component of equation (\ref{eq:V_w1}).

\noindent
{\bf Conduction current:}
Consider an ac potential $v_1^{ac}$ applied to contact 1, with contact
2 grounded. As a result, the potential of the well develops a time
dependent component, $v_w(\omega)$ (equation
(\ref{eq:V_w1})\cite{Comment1}). The linear 
response ac current flowing in contact $i \in 1,2$ consists of two 
terms (i) due to ac potential $v_1^{ac}$, with ac potentials in the 
well and contact 2 set equal to zero and (ii) due to ac potential 
$v_w(\omega)$ in the well, with ac potential in contacts 1 and 2 set 
equal to zero. The first component is $g_{i1} v_1^{ac}$. This follows
from the definition in equation (\ref{eq:g}). The second component is 
physically equivalent to setting the ac potential in the well equal 
to zero along with an ac potential $-v_w(\omega)$ applied to contacts
1 and 2.  The linear response 
current flowing in contact $i$ due to this component is
$-(g_{i1} + g_{i2}) v_w(\omega)$ (from equation (\ref{eq:g})). The
total ac conduction current flowing in contacts 1 and 2 is the sum of
the two components,
\begin{eqnarray}
\mbox{ac conduction current in contact 1}\! &=&\! 
                   g_{11}(\omega) (v^{ac}_1(\omega) - v_w(\omega))
                +  g_{12}(\omega) (- v_w(\omega)) \nonumber   \\
\mbox{ac conduction current in contact 2}\! &=&\! 
                   g_{21}(\omega) (v^{ac}_1(\omega) - v_w(\omega))
                +  g_{22}(\omega) (- v_w(\omega)) \mbox{ .}
\end{eqnarray}
Now, if external ac potentials $v^{ac}_1(\omega)$ and 
$v^{ac}_2(\omega)$ are applied to both contacts 1 and 2, then the 
conduction current in the two contacts is given by,
\begin{eqnarray}
I_1^c(\omega)\! &=&\! g_{11}(\omega) (v^{ac}_1(\omega) - v_w(\omega))
+  g_{12}(\omega) (v^{ac}_2(\omega) - v_w(\omega))\label{eq:I1c}   \\
I_2^c(\omega)\! &=&\! g_{21}(\omega) (v^{ac}_1(\omega) - v_w(\omega))
+  g_{22}(\omega) (v^{ac}_2 (\omega) - v_w(\omega)) \mbox{ .}
\label{eq:I2c}
\end{eqnarray}

The well charge is related to the conduction currents
by the continuity equation,
\begin{eqnarray}
\frac{dq(t)}{dt} = I_1^c (t)\; +\; I_2^c (t)\;\;\;===>\;\;\;  
q (\omega) = \frac{I_1^c (\omega)\; +\; I_2^c (\omega)}{- j \omega}
\mbox{ .} \label{eq:I^D1}
\end{eqnarray}
Substituting equation (\ref{eq:I^D1}) in equations (\ref{eq:V_w}) and
(\ref{eq:V_w1}), the ac potential of the device can be written as,
\begin{eqnarray}
v_w(\omega) = \alpha_2 \; v^{ac}_1(\omega)\; +\; \frac{I_1^c 
(\omega)\; +\; I_2^c (\omega)}{- j \omega C} \;\mbox{.} 
\label{eq:V_w2}
 \end{eqnarray}
Using equations (\ref{eq:I1c}), (\ref{eq:I2c}) and (\ref{eq:V_w2}),
the conduction currents in contacts 1 and 2 can be expressed in terms
of the g matrix elements:
\begin{eqnarray}
G_1^c(\omega) = \frac{I_1^c(\omega)}{v^{ac}_1} &=& + \frac{\alpha_1
g_{11} - \alpha_2 g_{12} + \frac{1}{-j \omega C} (g_{11} g_{22} -
g_{12} g_{21})}{1 + \frac{1}{-j \omega C} (g_{11} + g_{22} + g_{12}
+ g_{21})} \label{eq:I1cI} \mbox{,} \\
G_2^c(\omega) = \frac{I_2^c(\omega)}{v^{ac}_1} &=& - \frac{\alpha_2
g_{22} - \alpha_1 g_{21} + \frac{1}{-j \omega C} (g_{11} g_{22} -
g_{12} g_{21})}{1 + \frac{1}{-j \omega C} (g_{11} + g_{22} + g_{12}
+ g_{21})} \mbox{.} \label{eq:I2cI} 
\end{eqnarray}
Here $\alpha_1$ and $\alpha_2$ are fractional drops in the 
externally applied ac potential between the well and contacts 1
and 2 respectively, in absence of charge in the well.

\noindent
{\bf Displacement Current:}
The displacement current flowing in contact $\alpha$ consists of two
components. One component $\pm j \omega C$ (the plus and minus signs
are for the currents in the two different contacts) is due to the 
dielectric nature of the barrier and well. This component does not
depend on tunneling of charge from the contacts to well and
is not explicitly written in the remainder of the paper. The other
component, which is due to tunneling of charge from the contacts to 
well, is equal to $j \omega \frac{C_\alpha}{C}$. The total 
displacement current in the two contacts are given by,
\begin{eqnarray}
I_1^d(\omega) &=&  j \omega \frac{C_1}{C} q(\omega) 
               =  - \frac{C_1}{C} ( I_1^c (\omega) + 
                             I_2^c (\omega) )  \label{eq:I1d} \\
I_2^d(\omega) &=& j \omega \frac{C_2}{C} q(\omega) 
               =  - \frac{C_2}{C} ( I_1^c (\omega)  + 
			I_2^c (\omega) ) \mbox{ .} \label{eq:I2d}
\end{eqnarray}

Solving equations (\ref{eq:I1c})-(\ref{eq:I2d}), we get the following
expression for conductance contributions from conduction and 
displacement currents:
\begin{eqnarray}
G_\alpha^d(\omega) = \frac{I_\alpha^d(\omega)}{v^{ac}_1} =
\frac{C_\alpha}{C}\; \frac{\alpha_1 ( g_{11} + g_{21} ) - \alpha_2
( g_{22} + g_{12} )}{1 + \frac{1}{-j \omega C} (g_{11} + g_{22} +
g_{12} + g_{21})} \mbox{ ,} 
						 \label{eq:I1dI}
\end{eqnarray}
where, $\alpha = 1,2$ and $\omega$ has been suppressed from the
arguments of the g matrix elements.

\noindent
{\bf Total current: }
Using equations (\ref{eq:I1d}) and (\ref{eq:I2d}), it can be seen that
the total current (conduction plus displacement) in the contacts can 
be expressed in terms of the conduction currents,
\begin{eqnarray}
I_1(\omega) &=& I_1^{c}(\omega) + I_1^{d}(\omega) = \frac{C_2}{C} 
I_1^{c}(\omega) - \frac{C_1}{C} I_2^{c}(\omega) \label{eq:I1total} \\
\mbox{and } I_2(\omega) &=& I_2^{c}(\omega) + I_2^{d}(\omega) = 
\frac{C_1}{C} I_2^{c}(\omega) - \frac{C_2}{C} I_1^{c}(\omega) = 
		  	- I_1(\omega) \mbox{.} \label{eq:I2total}
\end{eqnarray}
We would like to emphasize that $I^c_1(\omega)$ and $I^c_2 (\omega)$
are the conduction currents calculated by including the contribution 
to ac potential in the well due to ac well charge density. In the 
remainder of the paper it is assumed that $\alpha_1$ and $\alpha_2$
are equal to fractional drops in the potentials across $C_1$ and $C_2$
(figure 1). Then, $\alpha_1 = \frac{C_2}{C}$ and 
$\alpha_2 = \frac{C_1}{C}$. Substituting equations (\ref{eq:I1cI}) and 
(\ref{eq:I2cI}) in equations (\ref{eq:I1total}) and (\ref{eq:I2total}),
the total ac conductance is:
\begin{eqnarray}
G_1 (\omega) &=& + \frac{{\alpha_1}^2 g_{11} + {\alpha_2}^2 g_{22} -
\alpha_2 \alpha_1 ( g_{12} + g_{21} ) + \frac{1}{-j \omega C} (g_{11}
g_{22} - g_{12} g_{21})}{1 + \frac{1}{-j \omega C} (g_{11} + g_{22} +
g_{12} + g_{21})} \label{eq:I1totalI}		\\
G_2 (\omega) &=& - G_1 (\omega) \mbox{.} 	\label{eq:I2totalI} 
\end{eqnarray}

\subsection{ \bf Calculation of the Conductance Matrix Elements: 
$g_{\alpha \beta}$}
\label{subsect:condmat}

The general expression for the ac conduction current in contact 
$\alpha$ taken from reference \onlinecite{Anantram1} are,
\begin{eqnarray}
&&i_{\alpha}(\omega) = \frac{e}{\hbar} \int_{-\infty}^{+\infty} 
\frac{dE}{2\pi} Tr \{ i_{\alpha}^{(1)}(E,\omega) +  
i_{\alpha}^{(2)}(E,\omega) +  i_{\alpha}^{(3)}(E,\omega) +  
i_{\alpha}^{(4)}(E,\omega) \}    \label{eq:linres} \\ 
&&\mbox{where, }   \nonumber  \\
&&i_{\alpha}^{(1)}(E,\omega) = \sigma_{\alpha}^{<}(E+\hbar \omega,E) 
[\;G^{r}(E+\hbar \omega) - G^{a}(E)\;]      \label{eq:linres_1}  \\ 
&& i_{\alpha}^{(2)}(E,\omega) = - j \Gamma_{\alpha}  
g^{<}(E+\hbar \omega,E)      \\
&&i_{\alpha}^{(3)}(E,\omega) =  g^{r}(E+\hbar \omega,E) 
\Sigma_{\alpha}^{<}(E)  -  g^{a}(E+\hbar \omega,E) 
\Sigma_{\alpha}^{<}(E+ \hbar \omega)  \\ 
&&i_{\alpha}^{(4)}(E,\omega) =  \sigma^{r}_{\alpha} 
(E+\hbar \omega,E) G^{<}(E)  -  G^{<}(E+ \hbar \omega) 
\sigma^{a}(E+\hbar \omega,E) \mbox{.} \label{eq:linres_4} 
\end{eqnarray}
Here the functions represented by capital letters are calculated in 
the steady state limit and the functions represented by small letters
are calculated to first order in the applied ac potential. $G^r$, 
$g^r$, $G^a$, and $g^a$ are retarded and advanced greens functions at
the site representing the well.  Similarly $\Sigma^r_\alpha$, 
$\sigma^r_\alpha$, $\Sigma^a_\alpha$, and $\sigma^a_\alpha$ are 
retarded and advanced self energies at the well site due to coupling 
with contact $\alpha$. The function $\Sigma_{\alpha}^{<}(E)$ represents
injection of electrons from contact $\alpha$ to device at energy E. 
$\sigma_{\alpha}^{<}(E+\hbar \omega,E)$ represents time-dependent 
injection from contact $\alpha$ to device.

Using the expressions for dc \onlinecite{Datta4,Mahan1} and ac 
\onlinecite{Anantram1} selfenergies, applying an ac potential to
contact $\beta$ yields,
\begin{eqnarray}
\Sigma^r_{\alpha}(E) &=& j \Gamma_\alpha (E) \label{eq:selfen1}\\
\Sigma^<_{\alpha}(E) &=&  j \Gamma_\alpha (E) f_\alpha(E)  \\
\sigma^r_{\alpha}(E+\omega , E) &=& \frac{\Sigma^r_{\alpha} (E) - 
\Sigma^r_{\alpha}(E + \omega)}{\omega} \delta_{\alpha \beta} \\
\sigma^<_{\alpha}(E+\omega , E) &=& \frac{\Sigma^<_{\alpha}(E) - 
\Sigma^<_{\alpha}(E + \omega)}{\omega} \delta_{\alpha \beta}
\;\mbox{,} \label{eq:selfen2}
\end{eqnarray}
where, $\alpha$ stands for contacts 1 and 2. In the expression for
retarded self energy, we only keep the imaginary part and neglect the
real part which represents a shift in the resonant 
energy~\cite{Mahan1}.

In the dc limit the expression for $\Gamma_\alpha (E)$ 
is~\cite{Lake1},
\begin{eqnarray}
\Gamma_\alpha (E) &=& \frac{w_{\alpha}^2}{w} \mbox{sin} (k_{\alpha } 
a) \;\;\; \mbox{ for } E 	\; > \; V_\alpha \label{eq:Gamma-up} \\
\Gamma_\alpha (E) &=& 0 \;\;\; \mbox{ for } E \; < \; V_\alpha 
\mbox{ . } \label{eq:Gamma-dn}
\end{eqnarray}
Due to absence of inter-mode
scattering, the greens functions take the following form (using 
expressions for $g$ and $G$ from references \onlinecite{Anantram1} 
and \cite{Datta4} respectively):
\begin{eqnarray}
G^r (E) &=& \frac{1}{E - \epsilon_{r0} + \Sigma (E)} 
\label{eq:Gr-phas.coh}		\\
G^< (E) &=& G^r (E) \Sigma^< (E) G^a (E) \\
g^r (E+\omega , E) &=& G^r (E+\omega) \sigma^r (E+\omega, E) G^r (E) 
\label{eq:grw-phas.coh}	\\
g^< (E+\omega , E) &=&  G^r (E+\omega) \sigma^<(E+\omega , E) G^a (E)
\;+\;  \nonumber \\
                   & & \;g^r (E+\omega, E) \Sigma^<(E) G^a (E) \;+\; 
\nonumber \\
		   & &  \hspace{-0.5in} G^r (E+\omega) \Sigma^< 
(E+\omega) g^a (E+\omega,E) \;\} \mbox{.}  \label{eq:g^<}
\end{eqnarray}
Using equations (\ref{eq:selfen1})-(\ref{eq:selfen2}) and equations 
(\ref{eq:Gr-phas.coh})-(\ref{eq:g^<}) in equation (\ref{eq:linres}), 
the various conductance matrix elements are calculated.

\section{\bf EFFECT OF CHARGING ON THE ac CURRENT}
\label{sect:charging}

Many references calculate the ac current by neglecting the effect of 
charging in the 
device~\cite{Frensley88,Buot93b,Jacoboni90,Liu91,Chen91,Fu93}. 
Specifically, reference \onlinecite{Chen91} calculates the ac
conduction currents flowing in the two contacts by neglecting the 
effect of charging. It is then asserted that the total ac current 
is equal to the average of calculated conduction currents in the 
two contacts:
\begin{eqnarray}
I(\omega) = \frac{1}{2} ( i_{1}^c(\omega) - i_{2}^c(\omega) ) 
\mbox{,} \label{eq:av} 
\end{eqnarray}
where, $i_{1}^c ({\omega})$ and $i_{2}^c ({\omega})$ are conduction
currents calculated by neglecting charging. 

To illustrate the importance of charging and to show that equation
(\ref{eq:av}) is valid only under a special circumstance, we summarize
our line of argument from the previous section:

\noindent
The total current is the sum of the conduction and displacement 
currents,
\begin{eqnarray}
I_1(\omega) &=& I_1^{c}(\omega) + I_1^{d}(\omega)  \nonumber  \\
\mbox{and } I_2(\omega) &=& I_2^{c}(\omega) + I_2^{d}(\omega) 
\mbox{.}
\nonumber
\end{eqnarray}
\noindent
The conduction currents here should be calculated by including the 
effect of the potential in the well due to  time varying charge
density in the well [equation (\ref{eq:V_w})]. The displacement 
currents are related to the time dependent charge density and are 
given by,
\begin{eqnarray}
I_1^d(\omega) &=&  j \omega \frac{C_1}{C} q(\omega)  \mbox{and}
\nonumber \\
I_2^d(\omega) &=& j \omega \frac{C_2} {C} q(\omega) \mbox{.}  
\nonumber
\end{eqnarray}
Now $q(\omega)$ in the above equations can be related to the
conduction currents using the continuity equation,
\begin{eqnarray}
\frac{dq(t)}{dt} = I_1^c (t)\; +\; I_2^c (t)\;\;\;===>\;\;\;
q (\omega) = \frac{I_1^c (\omega)\; +\; I_2^c (\omega)}{- j \omega}
\nonumber
\end{eqnarray}
Using the previous five equations, the total current can be expressed
in terms of the conduction currents (calculated by including the 
effect of charging in the well) as,
\begin{eqnarray}
I_1(\omega) &=& \frac{C_2}{C} I_1^{c}(\omega) - \frac{C_1}{C} 
I_2^{c}(\omega)  \nonumber \\
\mbox{and } I_2(\omega) &=& \frac{C_1}{C} I_2^{c}(\omega) - 
\frac{C_2}{C} I_1^{c}(\omega)  \mbox{.} \nonumber
\end{eqnarray}
From these equations, we see that equation (\ref{eq:av}) is an
appropriate expression for the conduction current only when both $C_1
= C_2$ and conduction currents are calculated by neglecting the effect
of charging. 

In terms of our notation involving the $g$ matrix elements, equation
(\ref{eq:av}) corresponds to the following equation which is obtained 
by setting  $C = \infty$ and $C_1 = C_2$ in equation 
(\ref{eq:I1totalI}):
\begin{eqnarray}
G_1(\omega) = -G_2(\omega) = \frac{1}{2} \; \{ \alpha_1 ( g_{11} - 
g_{21} )+ \alpha_2 (g_{22} - g_{12} ) \}\mbox{.}  \label{eq:WrongI}
\end{eqnarray}
In section \ref{subsect:num_1_mode}, we will numerically compare our
results obtained from equation (\ref{eq:I1totalI}) to those obtained 
from equation (\ref{eq:WrongI}).

\section{\bf Numerical Examples}
\label{subsect:num_1_mode}

In this section, we demonstrate the effect of charging on the ac 
conductance by comparing the conductances calculated with and without
charging. The values of $\epsilon_{r}$, $\Gamma_1$ and $\Gamma_2$ 
are chosen only to illustrate the discussion of section 
\ref{sect:charging}. Capacitances comparable to those used here are 
possible only in very narrow cross section devices. The discussion 
of section \ref{sect:charging} is however valid for broad area 
resonant tunneling devices as well, where the effect of
capacitances is equally important.
In Example 1, we start with a device which is symmetric both in the 
barrier strengths and capacitances, at zero bias. Here the 
conductance calculated from equations (\ref{eq:I1totalI}) and 
(\ref{eq:WrongI}) are comparable.  The effect of introducing an 
asymmetry in only the barriers (Example 2) and the capacitances 
(Example 3) of the structure in example 1 is then studied. In Example
4, we discuss the ac conductance of a device in the presence of an 
applied bias. In the numerical examples considered here, we have 
verified that the low frequency ac conductance is equal to the 
differential conductance of the dc I-V curve.

{\bf Example 1.} {\it Symmetric device at zero bias:}

The conductance versus frequency with and without charging [equations
(\ref{eq:I1totalI}) and (\ref{eq:WrongI})] are found to be the same 
(circles and crosses of figure 3). This is because in the limit
of a symmetric device at zero bias, the ac charge density in the well
is zero (This follows from equations (\ref{eq:I^D1}), (\ref{eq:I1cI}), 
(\ref{eq:I2cI}) and by noting that for a symmetric structure at zero 
bias $g_{ij}= g_{ji}$.). As a result both the ac potential and the 
displacement currents due to the charge in the well are zero. The 
parameters chosen in this example are: $V_{dc} = 0$V, the chemical 
potential of contact 1 ($\mu_1$) and contact 2 ($\mu_2$) are chosen to
be $10$meV, $\epsilon_{r} = 10$meV, $\Gamma_1 = \Gamma_2 = 0.1$meV at
$E = \epsilon_{r}$, $w = 2000$meV, $w_1 = w_2 = \frac{w}{16.7}$, $C_1 
= C_2 = 2 \times 10^{-16}$F and $kT = 0.015 meV$.

\noindent
{\bf Example 2.} {\it Effect of asymmetry only in the barrier 
strength:} 

The structure is identical to that in Example 1 except that
the barriers are asymmetric ($\Gamma_1 = 0.02$meV and $\Gamma_2 =
0.1$meV). Then, the results predicted by equations (\ref{eq:I1totalI})
and (\ref{eq:WrongI}) are comparable [figure 2(a)]. This can be 
explained as the total capacitance between the device and the contacts
is so large that the contribution to the ac potential in the well due 
to the $\frac{q(\omega)}{C}$ term in equation ({\ref{eq:V_w}) is 
negligible and that  $C_1 = C_2$ (see discussion in section 
\ref{sect:charging}).

In the case of a structure with smaller capacitances ($C_1 = C_2 = 
1 \times 10^{-16}$F), the results obtained from equations 
(\ref{eq:I1totalI}) and (\ref{eq:WrongI}) are different [figure 2(b)].
This is because when the total capacitance is small, the potential of 
the well is altered significantly by the charge in it. Then, the 
conduction currents calculated with and without charging included are
different. Note that equation (\ref{eq:WrongI}) does not predict a 
change in the ac conductance when the capacitances are changed without
altering the ratios $\alpha_1$ and $\alpha_2$ [see the circles and 
cross marks in figures 3(a) and (b)].

\noindent
{\bf Example 3.} {\it Effect of asymmetry only in the capacitances:}

A DBRTS where an asymmetry in the capacitances has been introduced in
example 1 ($C_1 = 1 \times 10^{-15}$F and $C_2 = 5 \times 10^{-15}$, 
the barriers are symmetric) is now considered. Here $C$ is large enough
that the $\frac{q}{C}$ term does not contribute substantially to the
ac potential in the well. The answers obtained from equations
(\ref{eq:I1totalI}) and (\ref{eq:WrongI}) are different (figure 3)
because unequal displacement currents flow in the two contacts when
$C_1 \neq C_2$. Then, a simple averaging
of the conduction currents as in equations (\ref{eq:av}) and
(\ref{eq:WrongI}) is no longer valid. Note that the answer for the
ac conductance from equation (\ref{eq:WrongI}) is identical in examples
1 and 3 because equation (\ref{eq:WrongI}) does not correctly account 
for the asymmetry in the capacitances.

\noindent
{\bf Example 4.} {\it ac conductance in the presence of an applied 
voltage:}

The dc bias is chosen to be $V = 5 mV$ for the example in figure 4 and
the self-consistently determined position of the resonance is 10.5362
meV. The values of the various parameters are $C_1=C_2= 1 \times
10^{-15}$F, $w = 2000$meV, $w_1 = \frac{w}{17.1}$, 
$w_2 = \frac{E}{17.1}$, 
$\epsilon_{r0} = 8.0$meV, $\mu_1=15$meV, $\mu_2=10$meV and $kT=
0.037$meV.  From figure 4(a), we see that the elements of
the g matrix, $g_{12}$ and $g_{22}$ are larger than $g_{21}$ and 
$g_{11}$ respectively. This feature can be understood by noting that 
the g matrix elements depend on the variation of the Fermi function 
in the contacts and that this variation is more rapid around the 
resonant energy in contact 2 than in contact 1 [equation
(\ref{eq:selfen2})].
Also, the real part of $g_{12}$ and the imaginary part of $g_{22}$ 
exhibit a peak around a frequency of 0.5meV because the resonant 
energy in the well is about 0.5362 meV above the chemical potential 
of contact 2. We are in the regime where the $\frac{q}{C}$ component 
of $V_w$ is negligible and $C_1=C_2$. So the ac conductance here 
agrees well with that obtained from equation (\ref{eq:WrongI}).  
From equation (\ref{eq:I1totalI}), we find that $G_1(\omega) = 
\frac{1}{4} ( g_{11} + g_{22} - g_{12} - g_{21})$. This expression 
explains why the ac conductance looks similar to $g_{22}$, the largest
of the g matrix elements [figure 4(b)]. 

On the other hand, we know from section \ref{sect:charging} that for 
a device where $C_1 \neq C_2$, the ac conductance depends on the 
ratio of $C_1$ and $C_2$. To illustrate this, we keep the values of 
the resonant energy, applied bias and all other parameters the same 
as those used in figure 4(b), except that $C_2 = 4 C_1 = 
4 \times 10^{-15}$F. From equation (\ref{eq:I1totalI}), $G_1(\omega)
= \frac{16}{25} g_{11} + \frac{1}{25} g_{22} - \frac{4}{25} 
( g_{12} + g_{21} )$. While the g matrix elements remain the same as 
in figure 4, it is obvious from the expression for $G_1(\omega)$ that
the conductance here is very different from the previous case 
[figure 4(c)]. In contrast, equation (\ref{eq:WrongI}) predicts the 
same value for the ac conductance in the two cases~\cite{footnote1}.

\noindent
{\bf Experiments:} We now make some remarks on the experimental 
conditions necessary to observe the differences in the ac conductance
discussed. Example 1 corresponds to a symmetric GaAs-AlGaAs structure
with identical barriers on either side. A structure with equal 
capacitances between the well and the two contacts but with different
coupling strengths to the contacts (example 2) can be constructed as 
follows. The coupling across the left barrier can be made weaker by 
increasing the barrier height. For AlGaAs, the dielectric constant does
not change significantly as the Al doping is increased in the left 
barrier so as to increase the barrier height. As a result, the barriers
will have nearly the same width and hence capacitances. With regards to
example 3, the capacitance across the second barrier can be made five 
times larger by making the barrier about five times thinner than the 
first barrier. To have similar transmission coefficients, the barrier
height of the second barrier should be correspondingly increased.
These requirements can probably be met with the present advances in
band gap engineering and the exact values of the barrier heights and
widths are easy to determine. What is more difficult to
achieve is the close proximity of the contacts to the barriers that we
have assumed in this paper. This assumption was however made only to
make the calculations simpler and more realistic calculations that are
beyond the scope of the present work can be carried out.

\section{\bf CONCLUSIONS}
\label{sect:conclusions}

\noindent
In this paper, we have calculated the ac conductance of a DBRTS by
including the effect of imaging of charge from the well to the two 
contacts and present useful expressions to calculate the ac 
conductance of a DBRTS. The formalism is applicable at high 
frequencies and in the presence of finite dc biases. 
The self-consistent inclusion of the effect of imaging of well
charge is central to calculating total currents which are equal in 
the two contacts. We find that including the effect of 
imaging of charge from the well to the contacts plays a significant 
role in determining the ac conductance of a DBRTS. The time varying
charge density in the well contributes to a flow of displacement 
currents equal to $\frac{C_1}{C} \frac{dq}{dt}$ and $\frac{C_2}{C} 
\frac{dq}{dt}$ in contacts 1 and 2 respectively ($q(t)$ is the well
charge). The strength of imaging which is modeled by the total
capacitance $C$ plays a role in determining the ac potential in the 
well via the $\frac{q(t)}{C}$ term. 
These features were illustrated using simple numerical examples in 
section \ref{subsect:num_1_mode}.
Some previous papers calculated the ac conductance of a DBRTS by the 
following procedure. The conduction currents across the two barriers 
are calculated by neglecting the contribution to the ac potential in 
the well due to imaging of the time dependent charge density. Then the
total ac current is taken to be the average of the conduction currents
flowing across the two barriers. In conclusion, we have shown that such
a procedure to calculate the ac conductance is correct only when  both 
the capacitance is symmetrical ($C_1 = C_2$) and the value of $C$ is 
large.

{\bf Acknowledgements: } I would like to thank Supriyo Datta 
for discussions on various aspects of transport in mesoscopic 
systems and David Janes for discussions pertaining to the
experimental feasibility of some of the results of the paper.
Part of this work was carried out at Purdue University and
financial support through Supriyo Datta's NSF grant (ECS-9201446-01)
is acknowledged.

\newpage

{\bf Figure Captions: }

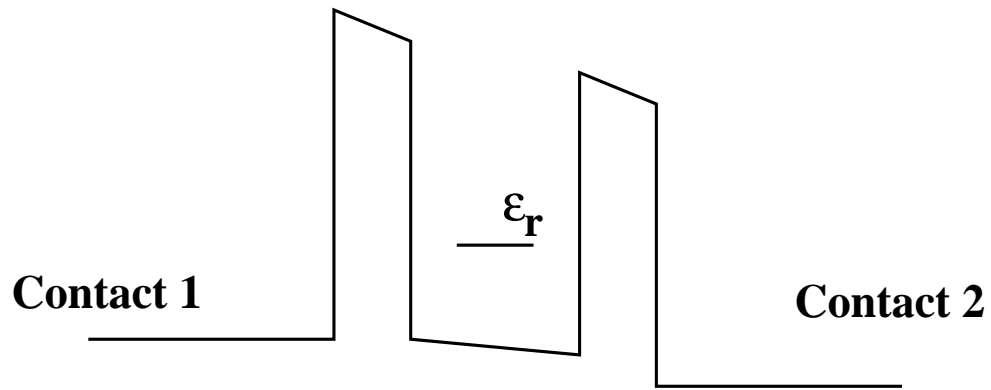
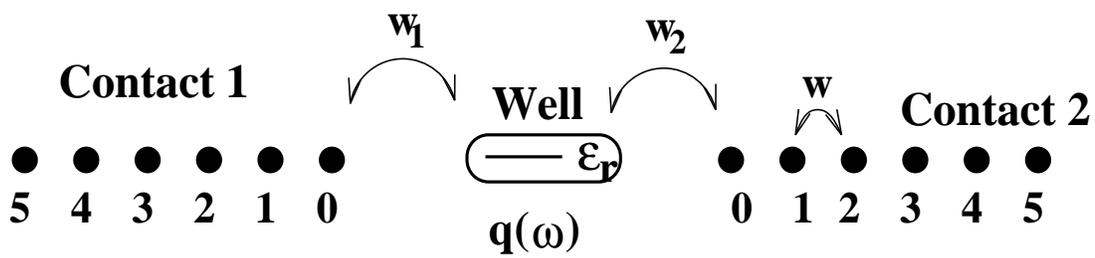
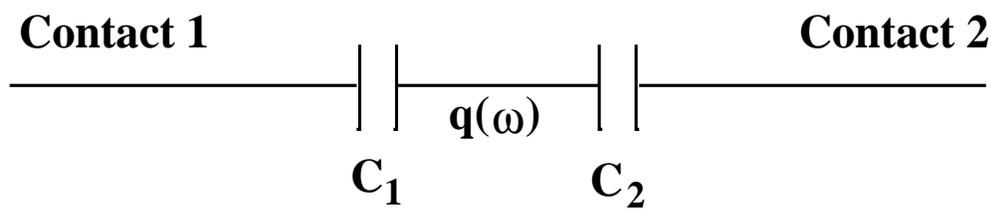
Figure 1:
(a) Band Structure of a DBRTS. (b) The DBRTS is modeled by the tight
binding Hamiltonian. (c) Imaging of charge from the well to the
contacts is accounted for by capacitive coupling between the well and
the contacts via capacitances $C_1$ and $C_2$.

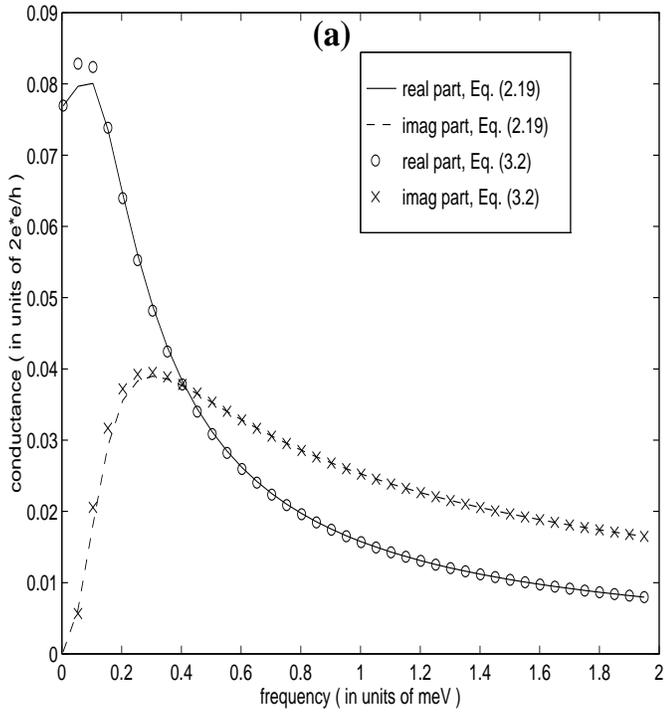
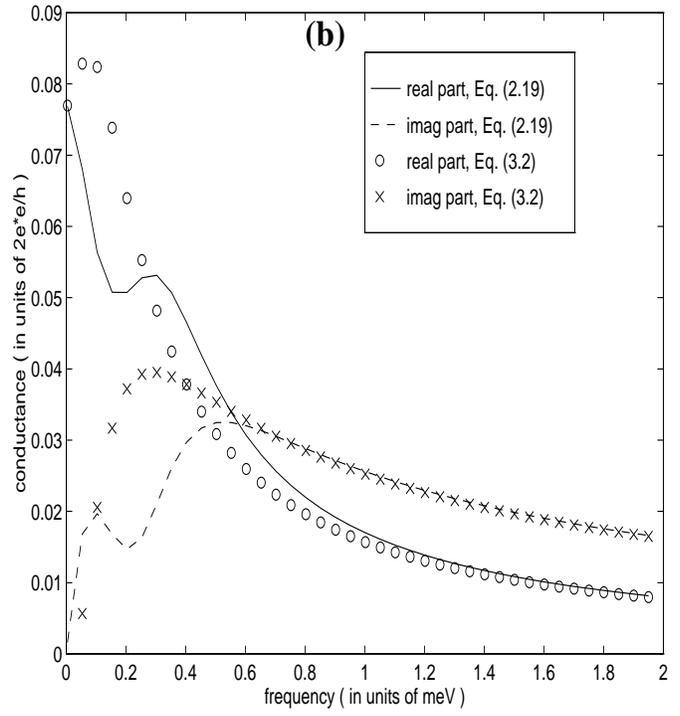
Figure 2:
{\it Asymmetry only in the barrier strengths:} {\bf (a)} barriers
are symmetric and the total capacitance is large. {\bf (b)} same
structure as (a) but the total capacitance is small.

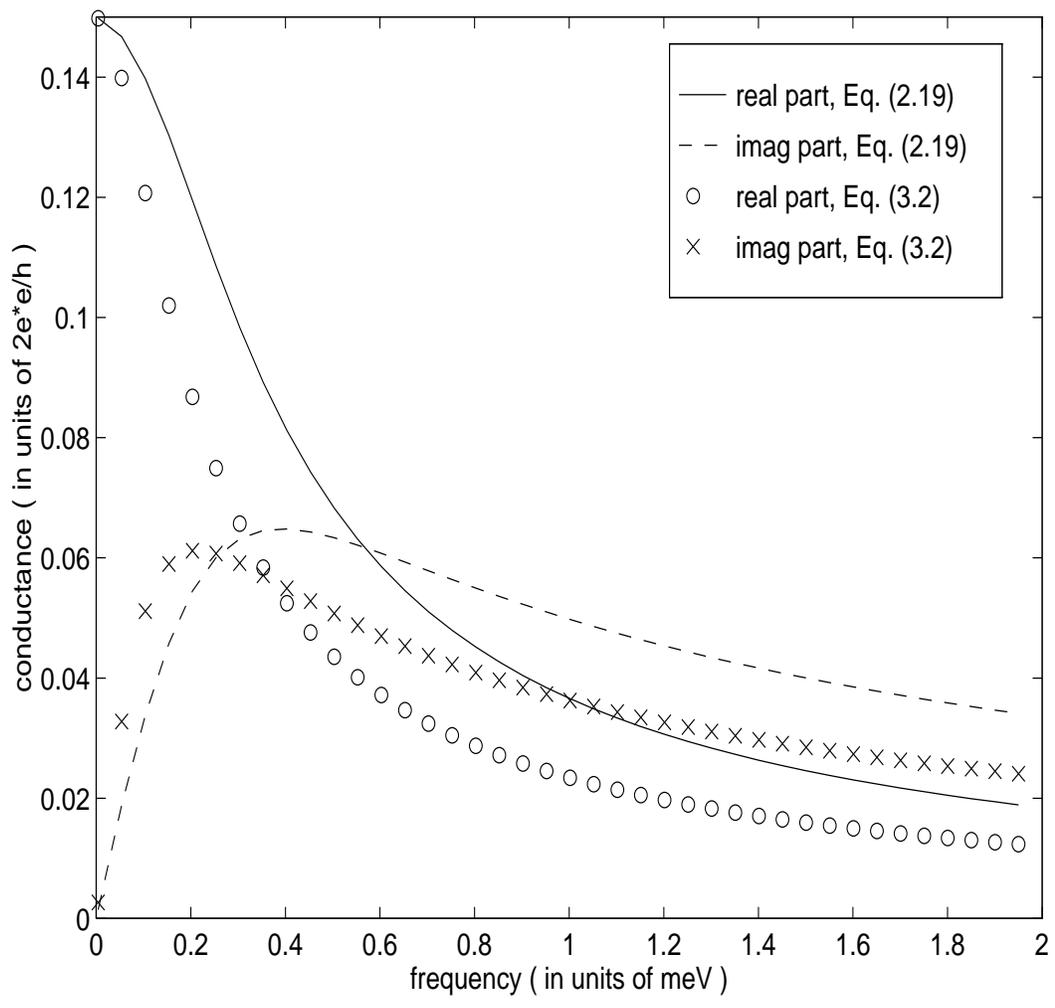
Figure 3:
{\it Asymmetry only in the Capacitances} 

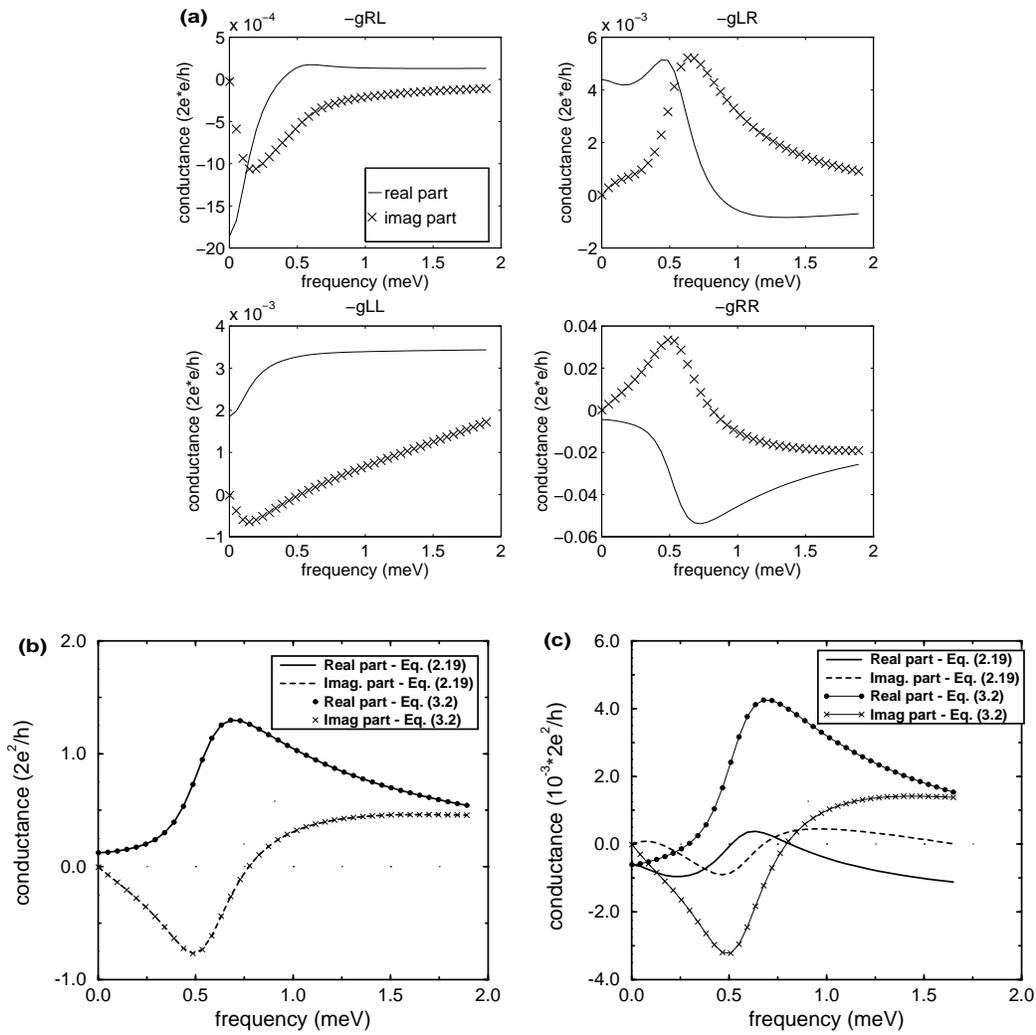
Figure 4:
{\it ac conductance in the presence of an applied bias.}  (a) The
conductance matrix elements. (b) The total conductance looks similar
to the largest conductance matrix element $g_{22}$ for this structure
when $C_1 = C_2$.  (c) The total conductance for a device where $C_2 =
4 C_1$ is however different from case (b).


\newpage


\begin{thebibliography}{10}

\bibitem{Buot93a}
F.~A. Buot, See Section 4 of Physics Reports, {\bf 234}, 73 (1993).

\bibitem{Frensley88}
W.~R. Frensley, Superlattices and Microstructures {\bf 4},  497  (1988).

\bibitem{Buot93b}
F.~A. Buot and A.~K. Rajagopal, Phys. Rev. B {\bf 48},  17217  (1993).

\bibitem{Jacoboni90}
C. Jacoboni and P.~J. Price, Solid State Comm. {\bf 75},  193  (1990).

\bibitem{Liu91}
H.~C. Liu, Phy. Rev. B {\bf 43},  12538  (1991).

\bibitem{Chen91}
L.~Y. Chen and C.~S. Ting, Phys. Rev. Lett. {\bf 64},  3159  (1990).

\bibitem{Fu93}
Y. Fu and S.~C. Dudley, Phys. Rev. Lett. {\bf 70},  65  (1993).

\bibitem{Landauer3}
R. Landauer, Physica Scripta {\bf T42}, 110   (1992).

\bibitem{Buttiker2}
M. B\"{u}ttiker, A. Pretre, and H. Thomas, Phys. Rev. Lett. {\bf 70},  
4114 (1993).

\bibitem{Buttiker3}
M. B\"{u}ttiker, H. Thomas, and A. Pretre, Phys. Lett. {\bf A180}, 
364, (1993).

\bibitem{Buttiker4}
M. B\"{u}ttiker, J. Phys. Condensed Matter {\bf 5}, 9361, (1993).

\bibitem{Anantram1}
M.~P. Anantram and S. Datta, Phy. Rev. B {\bf 51},  7632  (1994).

\bibitem{Christen1}
T. Christen and M. B\"{u}ttiker, Phys. Rev. {\bf B53} 2064, (1996).
This reference includes the effect of dephasing in the ac
conductance using voltage probes.

\bibitem{Sheard1}
F.~W. Sheard and G.~A. Toombs, Appl. Phys. Lett. {\bf 52},  1228
(1988).

\bibitem{Datta4}
S. Datta, {\em Electronic Transport in Mesoscopic Systems} (Cambridge
  University Press, Cambridge, United Kingdom, 1995).

\bibitem{Comment1}
From equations (\ref{eq:V_w}) and (\ref{eq:V_w1})),
$v_w(t) = v_w^{ac} (t) + \alpha_2 v_1^{ac} (t)$.

\bibitem{Mahan1}
G.~D. Mahan, {\em Many Particle Physics, Second edition} (Plenum 
Press, New York, 1991).

\bibitem{Lake1}
R. Lake, G. Klimeck, and S. Datta, Phys. Rev. B {\bf 47},  6427
(1993).

\bibitem{footnote1}
The position of the resonance is not calculated self-consistently here
as we only want to illustrate the effect of changing the capacitances.

\end{thebibliography}
\end{document}